# Pressure induced metallization with absence of structural transition in layered MoSe$_2$

Zhao Zhao[1,#,*], Haijun Zhang[2,1,#], Hongtao Yuan[3,4,#], Shibing Wang[4,5], Yu Lin[5], Qiaoshi Zeng[6,7], Gang Xu[3], Zhenxian Liu[8], G. K. Solanki[9], K. D. Patel[9], Yi Cui[3,4], Harold Y. Hwang[3,4], Wendy L. Mao[5,10]

*[1]Department of Physics, Stanford University, Stanford, CA 94305, USA*

*[2]National Laboratory of Solid State Microstructures and Department of Physics, Nanjing University, Nanjing 210093, China*

*[3]Geballe Laboratory for Advanced Materials, Stanford University, Stanford, CA 94305, USA*

*[4]Stanford Institute for Materials and Energy Sciences, SLAC National Accelerator Laboratory, Menlo Park, CA 94025, USA*

*[5]Department of Geological and Environmental Sciences, Stanford University, Stanford, CA 94305, USA*

*[6]HPSynC, Geophysical Laboratory, Carnegie Institution of Washington, Argonne, IL 60439, USA*

*[7]Center for High Pressure Science and Technology Advanced Research, Shanghai 201203, China*

*[8]Geophysical Laboratory, Carnegie Institution of Washington, Washington, D.C. 20015, USA*

*[9]Department of Physics, Sardar Patel University, Vallabh Vidyanagar, Gujarat 388120, India*

*[10]Photon Science, SLAC National Accelerator Laboratory, Menlo Park, CA 94025, USA*

[#] These authors contributed equally to this work.

*Corresponding authors: zhaozhao@stanford.edu*

**Layered transition-metal dichalcogenides have emerged as exciting material systems with atomically thin geometries and unique electronic properties. Pressure is a powerful tool for continuously tuning their crystal and electronic structures away from the pristine states. Here, we systematically investigated the pressurized behavior of $MoSe_2$ up to ~ 60 GPa using multiple experimental techniques and *ab initio* calculations. $MoSe_2$ evolves from an anisotropic two-dimensional layered network to a three-dimensional structure without a structural transition, which is a complete contrast to $MoS_2$. The role of the chalcogenide anions in stabilizing different layered patterns is underscored by our layer sliding calculations. $MoSe_2$ possesses highly tunable transport properties under pressure, determined by the gradual narrowing of its band-gap followed by metallization. The continuous tuning of its electronic structure and band-gap in the range of visible light to infrared suggest possible energy-variable optoelectronics applications in pressurized transition-metal dichalcogenides.**

Transition-metal dichalcogenides (TMDs) $2H-MX_2$ (M = Mo, W, and etc, X = S, Se, and Te) have recently attracted intense scientific and engineering interest because of their ease of fabrication and unique electronic structure[1–16]. TMDs have strong chemical bonding within each X-M-X trilayer and weak van der Waals (vdW) interaction between neighbor trilayers in their crystal structures. This quasi two-dimensional (2D) nature grants TMDs facile three-dimensional (3D) to 2D crossover through methods like exfoliation[1,2,4,5,8,9,12]. Band structure engineering on TMDs allows ones to explore exotic condensed matter phenomena and develop many potential applications. For example, the modification of their band structures from indirect-band-gap to direct-band-gap provides insights into opto-electronics and valley electronics[11,14–18]. So far, the electronic structure engineering of TMDs has mainly been achieved through the following experimental methods: i) applying electrical field to control the spin splitting and freedom of electrons[14,16], ii) utilizing quantum confinement with samples thinning down into monolayers of $MX_2$[11,15,17,18], and iii) employing stress or strain (by bending or stretching the thin films or employing substrates with different lattice constants)[19,20], as suggested by calculations[18,21–26].

Compared with the three methods mentioned above, high pressure is a powerful way to induce dramatic changes in their crystal structures and electronic structures[27,28]. This qualifies high pressure as a desirable approach to explore the tunability of TMDs. In particular, the ability to continuously tune the crystal and electronic structures away from the pristine states is crucial to a wide array of applications, e.g. electromechanical devices, energy-variable opto-electronics, and energy-variable photovoltaics. Various pressure-induced electronic evolutions such as insulator to semiconductor transitions or semiconductor to metal transitions have been observed in different materials[29–31]. However, many of these electronic transitions were accompanied by first-order structural transitions. A pressure-induced first-order structural transition, by definition, involves a discontinuous change in the volume (of the unit cell). And the corresponding discontinuity in its electronic structure could limit the energy tunability for potential opto-electronic or photovoltaic applications. To overcome this challenge, we need to discover TMDs with continuous structural and electronic response.

Previous high pressure studies on $MoS_2$ clearly demonstrated that a first-order structural transition ($2H_c$-$MoS_2$ to $2H_a$-$MoS_2$) occurred close to metallization[32–36]. This transition may relate to the vdW interactions in between neighbor S-Mo-S trilayers[36]. To prevent the $2H_c$ to $2H_a$ transition, the substitution of chalcogenide anions in $MoS_2$ is a potential route. Because $Se^{2-}$ has broader electron orbitals than $S^{2-}$'s that lead to stronger interlayer interactions, $MoSe_2$ may have totally different high pressure behavior. Experimentally, the structural and electronic behavior of compressed $MoSe_2$ remains to be fully explored[37–39]. Interestingly, recent calculations have predicted that $MoSe_2$ metallizes between 28 to 40 GPa while preserving the $2H_c$ structure[40].

In our work, high pressure up to ~ 60 GPa was generated by a diamond anvil cell (DAC). X-ray diffraction (XRD) data and Raman spectra data indicate that $MoSe_2$ preserves the $2H_c$ structure without any structural transition. *Ab initio* calculations for modeling the layer sliding process are presented to understand the contrasting behavior between $MoSe_2$ and $MoS_2$, and further predict the structural stability of $MoTe_2$. Infrared (IR) spectra data and temperature-dependent electrical resistivity data demonstrate the highly tunable transport properties of $MoSe_2$. Electronically, both experiments and calculations show that pressure strongly modulates its band structure from semiconducting to metallic.

## Results

**XRD and Raman spectroscopy.** The experimental set-up of the DAC is shown in Fig. 1a. Under compression, all XRD peaks for $MoSe_2$ continuously shift to larger $2\theta$ (smaller *d*-spacing) and no new peaks are observed (Supplementary Fig. 1 and Supplementary Note 1). Decompression of the sample shows the shifts of all peaks are reversible. All patterns are consistent with the $2H_c$ structure and representative Rietveld refinements are shown in Supplementary Fig. 2 and Supplementary Table 1. The absence of a first-order structural transition is further supported by equation of state (EOS) plots in Fig. 1b and normalized cell parameters vs pressure plots in Fig. 1c, as neither of them exhibits any discontinuity. To fit the pressure-volume relation, a third-order Birch-Murnaghan (BM) EOS is employed. With unit-cell volume $V_0$ fixed at the experimental value of 120.8 Å$^3$, the fitting yields a bulk modulus $B_0 = 62(1)$ GPa and a derivative of bulk modulus $B' = 5.6(1)$. The relatively large value of $B'$ suggests a strong change of volume compressibility under pressure.

At ambient conditions, the structure of $2H_c$-$MoSe_2$ features the X-M-X triple layers linked via vdW forces (Fig. 2a,b)[41,42]. During the initial compression, *c* direction is much more compressible than *a* direction due to the weak vdW interactions in between Se-Se layers along *c* (Fig. 1c), while higher pressure allows them to have nearly isotropic contractions. At the highest pressure ~ 60 GPa, *a* and *c* reduce by ~ 10 % and ~ 15 % respectively. The gradual closure of the vdW gap is suggested by tracking the ratio of Se-Mo layer distance to Se-Se layer distance, where it drops fast at low pressure but decreases much slower at high pressure (Supplementary Fig. 3). In addition, we measured the Raman spectra under pressure (Supplementary Fig. 4 and Supplementary Note 2). See Fig. 1d, the vibrational modes $A_{1g}$ and $E_{2g}$, and the spacing between them shift successively under pressure. These observations indicate that the crystal structure of $MoSe_2$ continuously evolves from a quasi 2D structure to an isotropic 3D one without a structural transition.

**Structural calculations.** Our XRD and Raman results appear to be surprising − At ambient conditions, $MoS_2$ and $MoSe_2$ are iso-structural in crystal structures and possess highly similar electronic structures, and it is therefore natural to assume that the $2H_c$ to $2H_a$ transition[32–35] would also occur in $MoSe_2$. Nonetheless, $2H_a$ structure does not fit the XRD patterns of $MoSe_2$ in the entire pressure region studied in this work. Though bearing highly similar Mo-Se chemical environments and the same space group ($P6_3/mmc$), $2H_a$ structure and $2H_c$ structure have distinct structural topologies. In $2H_a$ structure Mo occupies a 2b Wyckoff position while in $2H_c$ structure it occupies a 2c Wyckoff position. Also, the two adjacent units of X-Mo-X triple layers have contrasting stacking patterns in $2H_c$ and $2H_a$ (shown in Fig. 2a-d). To seek further support of the structural stability of $2H_c$-$MoSe_2$, we performed two sets of *ab initio* calculations. We first confirmed that $2H_c$-$MoSe_2$ is more stable than $2H_a$-$MoSe_2$, based on the experimental unit cell at the highest pressure ~ 60 GPa, consist with recent calculations showing the enthalpy difference between $2H_a$ and $2H_c$ of $MoSe_2$ keeps increasing from ambient pressure up to at least 130 GPa[40]. We then calculated the cell parameters at different volumes based on $2H_c$-$MoSe_2$ and the results agree well with our experimental data (Supplementary Fig. 5 and Supplementary Note 3). Note that the small discrepancy at low pressure comes from the limitations of *ab initio* calculations in describing the vdW interaction.

In order to understand the contrasting structural behavior between $MoS_2$[32–35] and $MoSe_2$[40], and see whether there is a predictable trend in TMDs, we further carried out layer sliding simulations for $MoS_2$, $MoSe_2$, and $MoTe_2$ at ~ 20 GPa. The side and top views of $2H_c$-type and $2H_a$-type structures are shown in Fig. 2a-d. The transition from $2H_c$ to $2H_a$ can be realized by systematically shifting half of the atoms (one unit of X-Mo-X triple layers) in a unit cell. As illustrated in Fig. 2a, we defined one sliding path by the red arrows for this transition. For $MoS_2$, the experimental unit-cell volume is fixed at the reported value at 20 GPa[35]. After initial relaxation of atomic positions within the $2H_c$-type structure, the S-Mo-S layer distance is set to be a constant during the layer sliding. The same procedures are performed for $MoSe_2$ at 23 GPa (using our experimental data) and $MoTe_2$ at 20 GPa (based on previous theoretical data[40]). The total energies of $2H_c$ are set to be zero as the references. Fig. 2e shows the relative energies as a function of the relative layer sliding, i.e. 0 represents $2H_c$ and 1 represents $2H_a$. $MoX_2$ needs to overcome an energy barrier in order to undergo the $2H_c$ to $2H_a$ transition. The maximum energy barrier is ~ 0.3 eV for $MoSe_2$ and ~ 0.5 eV for $MoTe_2$, which are significantly higher than ~ 0.15 eV for $MoS_2$. More importantly, $2H_a$-$MoSe_2$ and $2H_a$-$MoTe_2$ bear higher energies than the initial $2H_c$ structures, which would make this transition unfavorable. However, in the case of $MoS_2$, the $2H_a$ structure becomes energetically favorable. Note that the X-Mo-X layer distance is fixed in this set of calculations. Realistically in compressed $MoS_2$, the S-Mo-S distance and unit cell volume drops during the $2H_c$ to $2H_a$ transition[32–35], which allows the total change in enthalpy to be continuous at zero temperature.

**IR spectroscopy.** The lattice response of $MoSe_2$ at high pressure will inevitably change its electronic structure, and thus its optical properties which strongly depend on the electronic structure. Our data shows that MoSe$_2$ undergoes a large electronic evolution where band-gap narrowing is followed by metallization of MoSe$_2$. Fig. 3 shows the measured synchrotron IR spectra and its analysis (details are shown in Supplementary Note 4). Below 16.3 GPa, the transmittance spectra (Fig. 3a) look similar, where a transmission window extends from 0.06 eV to 1.0 eV. At pressure above 20.2 GPa, the 0.3-1.0 eV region develops into a tilted curve and keeps collapsing into lower energy region, indicating the band-gap's narrowing. At above 40.7 GPa to the highest pressure, nearly zero transmission is observed in between 0.15 to 1.0 eV, suggesting the metallization of MoSe$_2$. Another way to interpret the IR data is from the viewpoint of the optical density (OD) $A_\lambda$ (see Supplementary Fig. 6 for the plot of OD versus energy). OD or $A_\lambda$ is defined as $-\log T$ ($T$ as the transmittance). It can be easily seen from the energy-pressure-OD map in Fig. 3b that a clear changeover of low OD (in semiconducting state) to high OD (in metallic state) occurs between 20 to 35 GPa.

For an indirect-band-gap semiconductor, the absorption coefficient is proportional to the square of the photon energy and band gap[43]. Using this empirical model for semiconductors, we obtained the indirect-band-gap $E_g$ via linear extrapolations of $(h\upsilon A_\lambda)^{1/2}$ where $h\upsilon$ is the energy of incident light. A representative extrapolation is shown in Supplementary Fig. 7. The fitted $E_g$ value at 20.2 GPa is 0.4 eV. From 20.2 to 35.1 GPa, $E_g$ keeps decreasing (see Fig. 3c). From 40.7 GPa to the highest pressure, $E_g$ is nominally zero. We notice that the trend of band-gap decrease at below 41 GPa could not be well described by a linear fitting. The non-linearity is also shown in previous calculations on band-gap's dependence on applied strain[23]. The lack of data points and inaccuracies in optical measurement prevent us from determining the best function for the band-gap − pressure relation. However, as a simple approach to guide eyes, we fit the data with a parabolic curve, which yields $E_g = -0.08(2)\ P + 0.0010(3)\ P^2$, indirect band gap ($E_g$) in unit of eV and pressure ($P$) in unit of GPa. The extrapolated band-gap at ambient pressure is 1.6(3) eV, which is in good agreement with previous reports[18,44,45].

**Electrical resistivity.** We also measured the temperature-dependent resistivity up to ~ 42 GPa (Supplementary Note 5 and Supplementary Fig. 8). At low pressures (Fig. 4a), the temperature ($T$) - resistivity ($\rho$) curves at below 23.4 GPa exhibit negative $d\rho/dT$ throughout all temperatures, indicating the presence of a semiconducting state. From 27.0 GPa to 37.0 GPa, the high-temperature region shows positive $d\rho/dT$ whereas the low-temperature region has negative $d\rho/dT$ (see Fig. 4b). At above 41.0 GPa, positive $d\rho/dT$ is observed in all temperatures, implying the metallization of MoSe$_2$. A comparison of our room temperature resistivity data on MoSe$_2$ (Fig. 4c) with previous data on MoS$_2$ shows that there are dissimilar trends in between them[34,35]. For MoS$_2$, the room temperature resistivity dropped strongly at below ~ 15 GPa and then reached a plateau at higher pressure[34,35], which was related to a first-order structural transition. In sharp contrast, for MoSe$_2$, the decrease of its room temperature resistivity is nearly exponential, fit by log ($\rho$) = 1.9(1) − 0.134(5) $P$, resistivity ($\rho$) in unit of Ω cm$^{-1}$ and pressure (P) in unit of GPa. Pressure allows the electrical resistivity of MoSe$_2$ to decrease more than six orders of magnitude from ambient to 41.6 GPa.

**Electronic structure.** To better understand the electronic evolution of MoSe$_2$ that determines its highly tunable optical and electrical transport properties, we performed *ab initio* calculations on the electronic structure of MoSe$_2$ at four representative pressures. The results undoubtedly demonstrate the pressure-induced band-gap narrowing and metallization. At ambient pressure, seen from Fig. 5a, the band structure is consistent with previous results[18,44,45]. It has a direct-band-gap $E_{K-K}$ (~1.8 eV) at K and an indirect-band-gap (~1.3 eV) that locates in between Γ and Γ-K conduction band (CB) valley. The bottom of CB between Γ and K mainly origins from the Mo *dxy* and *dx$^2$ - y$^2$* orbitals, and the top of valence bands (VBs) at Γ comes from the Mo *dz$^2$* orbital. Meanwhile, the *dxz* and *dyz* dominated CBs are further above from the Fermi-level ($E_F$). Higher pressure results in a strong decrease of its indirect band gap and induces large movements of the orbitals towards the $E_F$. At 23 GPa, shown in Fig. 5b, the indirect band gap becomes as small as 0.5 eV. Albert decreasing, the direct-band-gaps remain at large values, e.g. $E_{K\text{-}K}$ is ~ 1.4 eV. Remarkably, pressure allows the *dxz* and *dyz* orbitals to gain more overlap with Se *p* orbitals and thus largely widen their band dispersions. In comparison, the *dxy* and $dx^2$–$y^2$ orbitals are less impacted due to smaller overlap with Se *p* orbitals. As a consequence (see Supplementary Note 6), one *dxz* and *dyz* dominated CB quickly goes down at K point and forms two CB valley minimum together with the previous *dxy* and $dx^2$ - $y^2$ dominated CB.

Metallic band structures are obtained by further increasing the pressure. For example, seen from Fig. 5c at 41 GPa, there lies density of states across the $E_F$. The *dxy* and $dx^2$ - $y^2$ dominated CB valley minimum crosses below the $E_F$, while the other CB valley minimum is still slightly above the $E_F$ (see Supplementary Fig. 9). At 58 GPa, shown by Fig. 5d, both CB valley minima cross below the $E_F$. It is worth mentioning that even at as high as 58 GPa, the "indirect" feature of the band structure is still well maintained: although the CBs and VBs overlap in energy range, no direct cross is seen. To be specific, the energy separation at K is as large as ~ 0.6 eV. Meanwhile, the relative shifts of CBs and VBs generate a number of hole pockets (e.g. at Γ and A) and electron pockets (e.g. at K). These pockets may largely affect the low-temperature electrical and optical transport properties of $MoSe_2$.

## Discussions

Our experiments and calculations clearly demonstrate the absence of structural transition in $MoSe_2$. One empirical understanding of the contrasting behavior in $MoS_2$[32–36] and $MoSe_2$ involves the different localizations of *p* orbitals among chalcogenide anions $S^{2-}$, $Se^{2-}$, and $Te^{2-}$. The 3*p* orbitals of $S^{2-}$ dominate the electronic structure in $MoS_2$ while the 4*p* and 5*p* orbitals are primary for $MoSe_2$ and $MoTe_2$ correspondingly. 4*p* and 5*p* orbitals are much more delocalized than 3*p* orbitals, which would give rise to strong interaction within the vdW gaps of MoSe$_2$ and MoTe$_2$ to prevent this sliding process, vice versa for MoS$_2$. We can safely conclude that it is easier for 2H$_c$-MoS$_2$ to experience a structural transition through sliding in between neighbor S-Mo-S layers, but this does not apply to MoSe$_2$ or MoTe$_2$. Beside from the effects of chalcogenides anions, the effects of transition metal cation should also be considered in determining the stabilities of layered structures. It is worth noting that recent calculations proposed that the interlayer Mo-Mo *d*-electron propagation should also be considered in determining the stability of layered structures[40]. Cation substation is also expected to change the interlayer interactions. Previous studies on WS$_2$ and WSe$_2$[46–48] showed that they experience continuous lattice contractions under pressure. W$^{2+}$ has broader electron orbitals and may introduce stronger interlayer interactions than Mo$^{2+}$, which results in the absence of layer sliding in WS$_2$. In a more general perspective, stronger interlayer interactions help stabilize the structures of TMDs and are more likely to yield continuous lattice response under extreme environments such as pressure.

Previous studies have reported many electronic transitions such as insulator to metal or semiconductor to metal transitions in the group of binary chalcogenides, see e.g. Bi$_2$X$_3$[31,49,50], Sb$_2$X$_3$[51–53], and Ag$_2$X[54–56]. For structures starting with vdW gaps at ambient conditions, the closure of their vdW gaps is generally accompanied or followed by first-order structural transitions where large structural re-constructions or atomic movements take place[43–48]. However, in the case of MoSe$_2$, the metallization process does not involve any sudden change in the crystal structure, which allows its electronic structure to be continuously tuned. Through multiple experimental techniques combined with *ab initio* calculations, we demonstrate that the band-gap of MoSe$_2$ (in the range of visible to IR region) has a strong dependence on pressure. This may allow MoSe$_2$, one representative TMD, to be applied in energy-variable opto-electronics and photovoltaics, although the limitation of sample size (0.05 to 0.1 mm) must be taken into account in future investigation.

Compared with others methods of band structure engineering approachable by experiments[11–18], pressure is the only way to metalize $MoSe_2$ and $MoS_2$[34,35]. This highlights pressure's dramatic role in tuning the electronic properties of TMDs. Our different aspects of the experimental data show good agreement with the recent calculations on $MoSe_2$[40]. This further suggests $2H_c$-$MoSe_2$ a suitable material with a concise unit cell for testing and improving first-principles calculations that can be probed by experiments. If future experiments, i.e. by applying non-uniaxial compression, are able break the inversion symmetry of the crystal structure of pressurized $MoSe_2$, the spin and valley electronics of $MoSe_2$ would be largely different when we consider the large shifts of the CBs and VBs[5,12,14,16]. More importantly, the novel scenario of excitonic insulator may be experimentally realizable in $MoSe_2$[40], while in $MoS_2$ the complexity of the layer sliding structural transition may prevent the formation of this electronic state[32–36]. For the pressurized metallic $MoSe_2$, its distinguished "indirect" band structures and electronics state filled with electron holes and pockets await further exploitation in condensed matters physics, i.e. charge density wave or superconductivity may be found in TMDs at higher pressure[35,36,40].

In conclusion, we comprehensively studied the high pressure behavior of $MoSe_2$ up to ~ 60 GP through a series of structural, vibrational, optical, and electrical measurements combined with *ab initio* calculations. $2H_c$-$MoSe_2$ evolves from an anisotropic 2D layered structure to an isotropic 3D one without any sudden structural change under pressure. Our layer sliding calculations highlight the role of the chalcogenide anions in stabilizing either $2H_a$ or $2H_c$ layered patterns. Electronically, $MoSe_2$ undergoes a semiconductor to metal transition, and correspondingly exhibits highly tunable optical and electrical properties. Upon compression, the "indirect" feature of its electronic structure is robustly conserved with the appearance of two conduction band minima. The large and continuous tuning of its electronic structure may have potential applications in energy-variable (visible to IR) opto-electronics and photovoltaics.

## Methods

### Sample growth

High quality stoichiometric MoSe$_2$ single crystals were grown by direct vapor transport technique[39,44,57]. Elemental Mo and Se (99.9% purity, purchased from *Koch Light Ltd.*) of the stoichiometric ratio were sealed in a quartz ampoule at pressure better than 10$^{-5}$ Torr. The ampoule was placed in a two-zone horizontal furnace where the temperatures were slowly raised from room temperature to 1060 ˚C and 1080 ˚C for growth zone and source zone respectively. This temperature gradient was then maintained for ~ 188 hours to produce to single crystal platelets of MoSe$_2$. The shiny and gray crystals have a typical thickness of ~ 4 μm and area of ~ 3mm ×3mm. The purity and homogeneity are checked by electron microprobe analysis.

### High pressure experiments

Single crystals of MoSe$_2$ were used for the high pressure IR, Raman, and resistivity measurements. Powders of MoSe$_2$ were grounded from single crystals for the high pressure XRD measurements. Ruby spheres were used for determining pressure for all experiments. Neon was used as the pressure transmitting medium for the XRD and Raman experiments. Angle dispersive XRD experiments were performed at beamline 16-BMD of the Advanced Photon Source (APS), Argonne National Laboratory (ANL). The Rietveld fitting was performed using GSAS-EXPGUI package[58]. The Raman spectra were collected using a Renishaw inVia micro Raman system with a 514 nm laser excitation line. High-pressure IR measurements were conducted in beamline U2A of the National Synchrotron Light Source (NSLS), Brookhaven National Laboratory (BNL). A MoSe$_2$ single crystal was sandwiched between the pressure transmitting medium (KBr) and one side of the culet. Infrared microspectroscopy was performed on a Bruker Vertex 80v FT-IR spectrometer coupled to a Hyperion-2000 microscope with a MCT mid-band detector. Fringes in raw IR data were removed by filtering high frequency terms after

Fourier transformation. For temperature-dependent four-probe resistivity measurement, cubic BN was used as the insulating layer, and four electrodes were cut from Pt foils. The temperature-dependent sheet resistance of the sample was measured with Van der Paul geometry by cooling down to 10 K in a liquid helium cryostat. Pressures were measured at room temperature. More details are described in the supplementary information.

## *Ab initio* calculations

The Vienna *ab initio* Simulation Package[59,60] was employed to optimize crystal structures and calculate electronic structures with the framework of local density approximation density functional theory[61]. The projector augmented wave[62] pseudo-potential was used and the kinetic energy cutoff was fixed to 450 eV for all the calculations. For structural calculations in comparison with experiments, we fixed the unit cell volume and fully relaxed lattice constants and atoms. For band structure calculations, the lattice constants were fixed to be experimental values, and then fully relaxed atoms. HSE06 hybrid function[63] was employed to improve the band structure calculations.

**Author information** * zhaozhao@stanford.edu

**Acknowledgments**

We thank Y. Xu at Stanford University for useful discussions. We thank S. Tkachev, C. Park, S. Sinogeikin, H. Yan, and Y. Meng at APS for their technical assistance. Z.Z., H.T.Y., Y.L., H.Y.H., Y.C., and W.L.M. are supported by the Department of Energy (DOE), Basic Energy Sciences (BES), Materials Sciences and Engineering Division, under Contract DE-AC02-76SF00515. H.Z. is supported by ARO W911NF-09-1-0508. K.D.P. and G.K.S. are funded by a major research project from University Grants Commission, New Delhi, India. HPCAT operations are supported by DOE-NNSA DE-NA0001974 and DOE-BES, DE-FG02-99ER45775, with partial instrumentation funding by NSF MRI-1126249. APS is supported by DOE-BES, DE-AC02-06CH11357. U2A is supported by COMPRES under NSF Cooperative Agreement EAR 11-57758 and DOE-NNSA DE-FC03-03N00144, CDAC. NSLS is supported by DOE-BES, DE-AC02-98CH10886.

**Author contributions**

Z.Z., H.Z., H.T.Y., and W.L.M. designed the project. G.K.S. and K.D.P. synthesized the samples. Z.Z., H.T.Y., S.W., Y.L., Q.Z., and Z. L. conducted the experiments. H.Z. performed the calculations. Y.C., H.Y.H, and W.L.M. supervised the project and all authors contributed to data discussions. Z.Z., H.Z., and H.T.Y., wrote the paper with inputs from all others.

**Figures:**

**Figure 1**

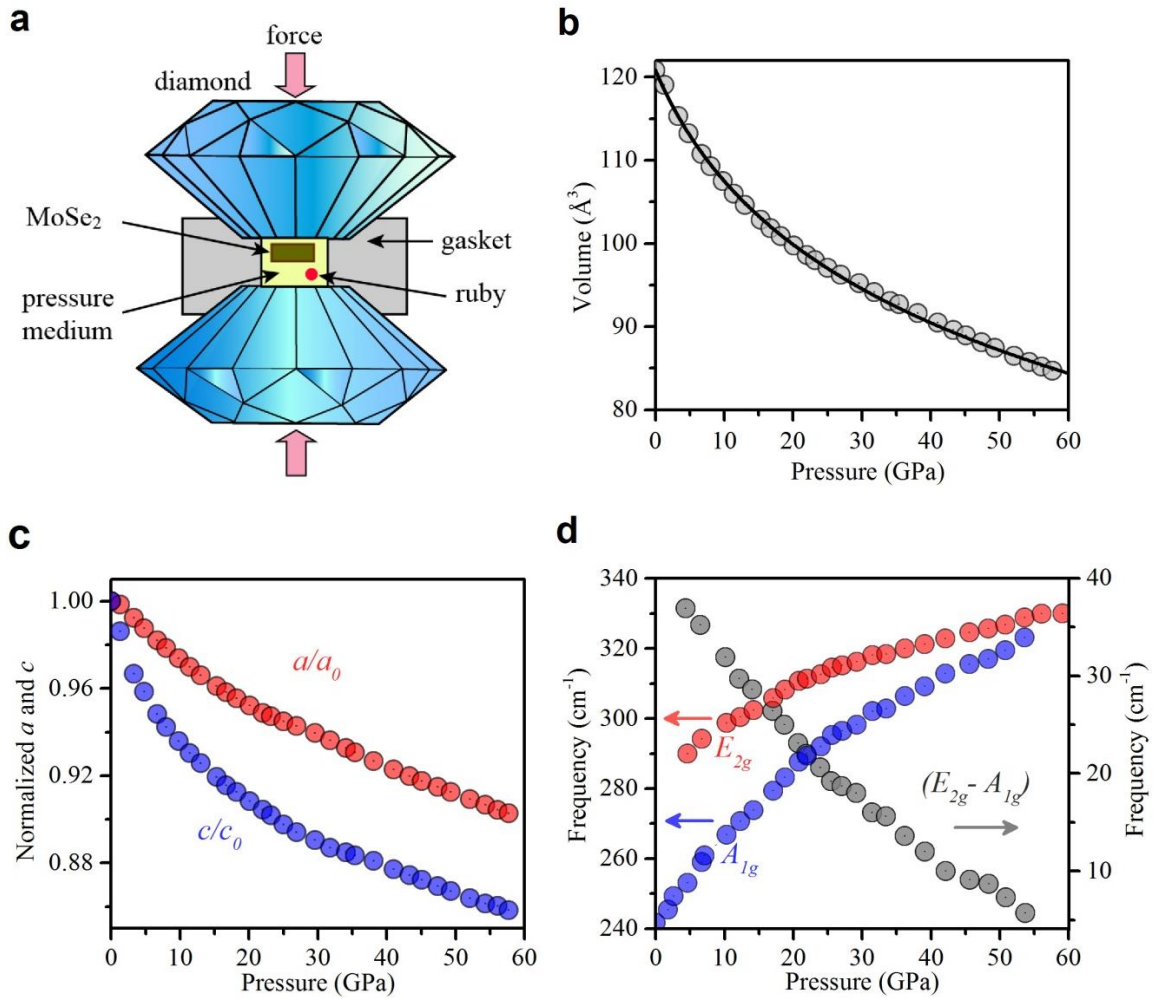

**Figure 1 | Experimental set up, and structural and vibrational responses under pressure.** (**a**) Schematic of the high pressure DAC set up. (**b**) Pressure-volume data from XRD measurement and the curve represents a third-order BM-EOS fitting. (**c**) Normalized cell parameters $a/a_0$ and $c/c_0$ versus pressure. The error bars given by EXPGUI-GSAS are smaller than the size of the markers. (**d**) Evolution of vibrational modes $A_{1g}$ and $E_{2g}$, and their difference ($E_{2g} - A_{1g}$) under pressure, measured by Raman spectroscopy.

**Figure 2**

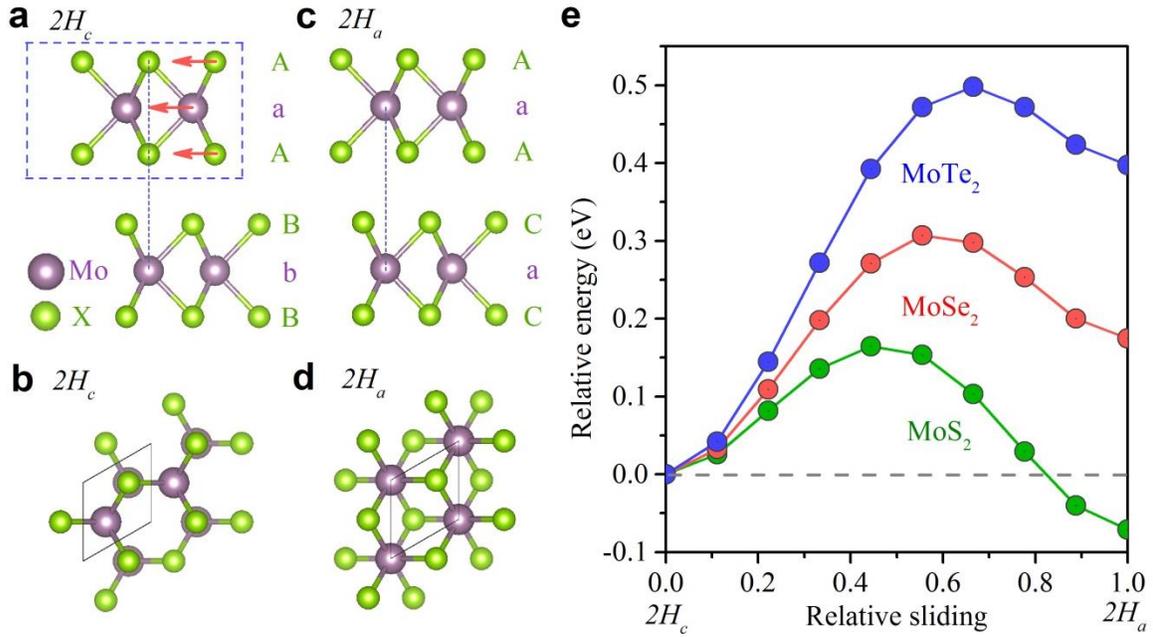

**Figure 2 | Structural details of $2H_c-$ and $2H_a-MoX_2$, and layer sliding calculations.** (**a**) The side view (projected on *ac* plane) of $2H_c$ structure in $MoS_2$, $MoSe_2$, and $MoTe_2$. X represents S, Se, and Te. The red arrows represent one sliding path for the $2H_c$ to $2H_a$ transition, where one unit of X-Mo-X triple layers (marked by a blue box) shifts systematically in *ab* plane. (**b**) The top view (projected on *ab* plane) of $2H_c$ structure. (**d**) The side view of $2H_a$ structure. (**c**) The top view of $2H_a$ structure. (**e**) The total energy of $MoS_2$, $MoSe_2$, and $MoTe_2$ as a function of relative sliding from $2H_c$ to $2H_a$. The total energies of $2H_c$ structures are set to be zero as references, marked by a broken line. For $MoS_2$, the unit cell volume was fixed at experimental data under 20 GPa[35]. After electronic relaxation of atomic positions, the S-Mo-S layer distance of structure was set to be a constant during the layer sliding. The same procedures were performed on $MoSe_2$ under 23 GPa (our experimental data) and $MoTe_2$ under 20 GPa (theoretical data[40]).

**Figure 3**

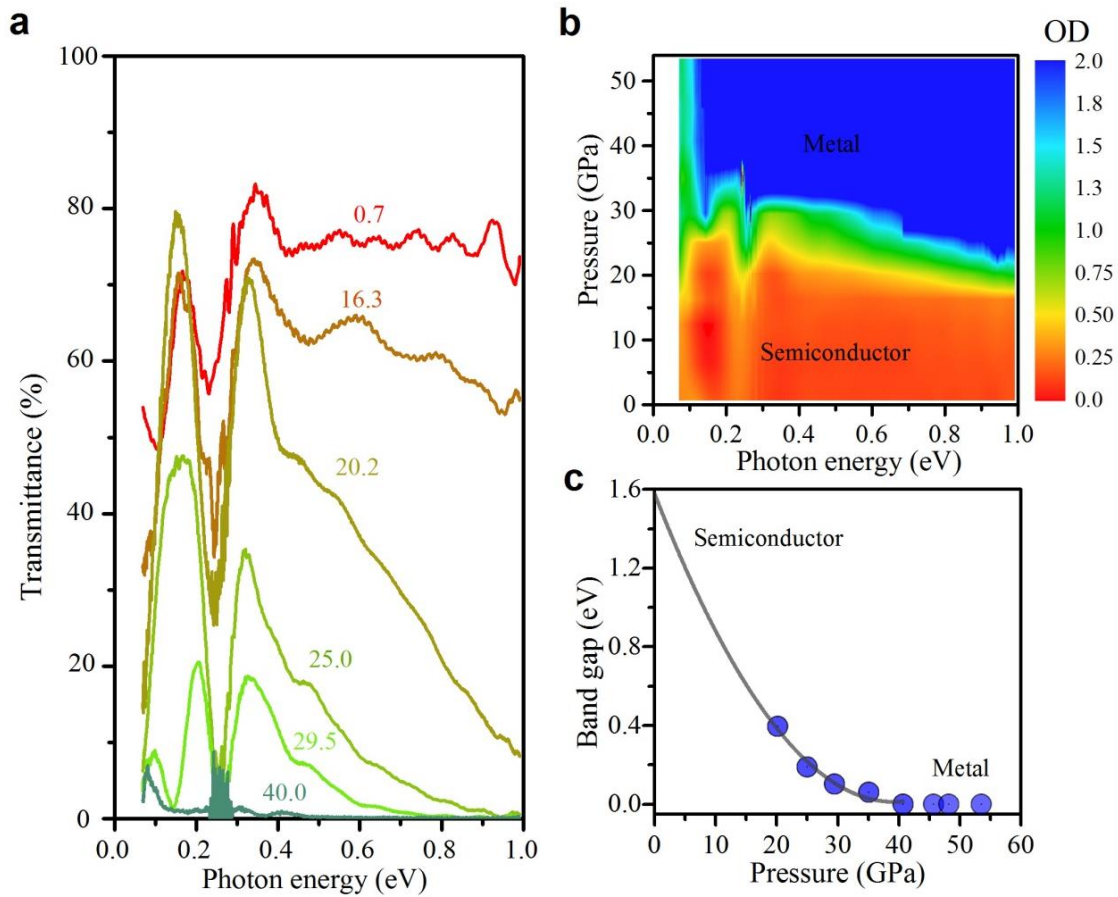

**Figure 3 | IR transmittance spectra and analysis.** (**a**) Representative IR transmittance spectra at high pressures, numbers show pressures in unit of GPa. The 0.23-0.28 eV region is obscured by diamond phonon absorptions from the DAC. (**b**) Pressure−energy−optical density (OD) contour, OD is defined as –log(*T*) while *T* is the transmittance. (**c**) Evolution of band gap under pressure. Circles are extrapolated indirect band gaps and the curve shows a parabolic fit of the band gap versus pressure. The band gap closure is observed at ~ 40 GPa.

**Figure 4**

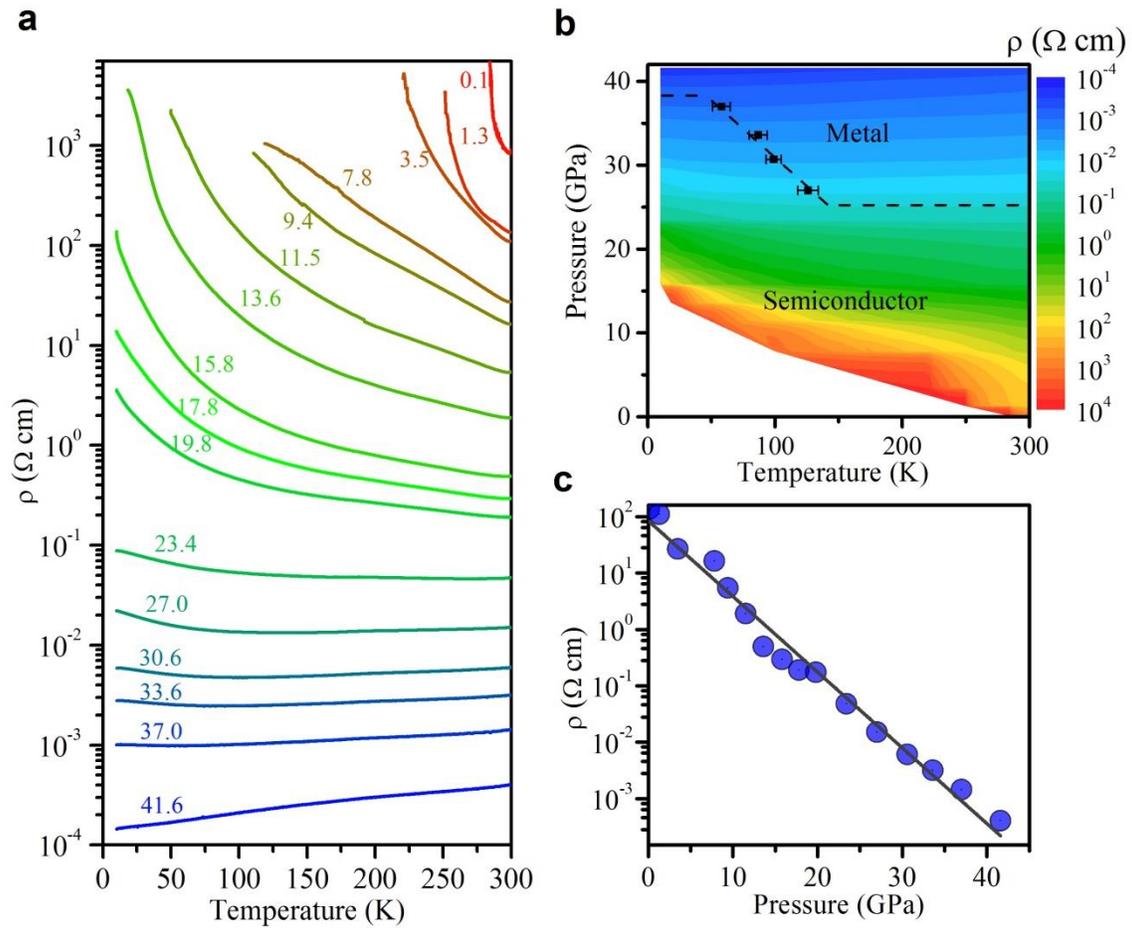

**Figure 4│Temperature-dependent resistivity data and analysis.** (**a**) Temperature-resistivity curves at different pressure, numbers show pressures in GPa. (**b**) Temperature-pressure-resistivity contour map. (**c**) Room temperature resistivity versus pressure, the line shows a linear fitting of log $\rho$ vs pressure (equivalent for an exponential fitting of $\rho$ vs pressure).

**Figure 5**

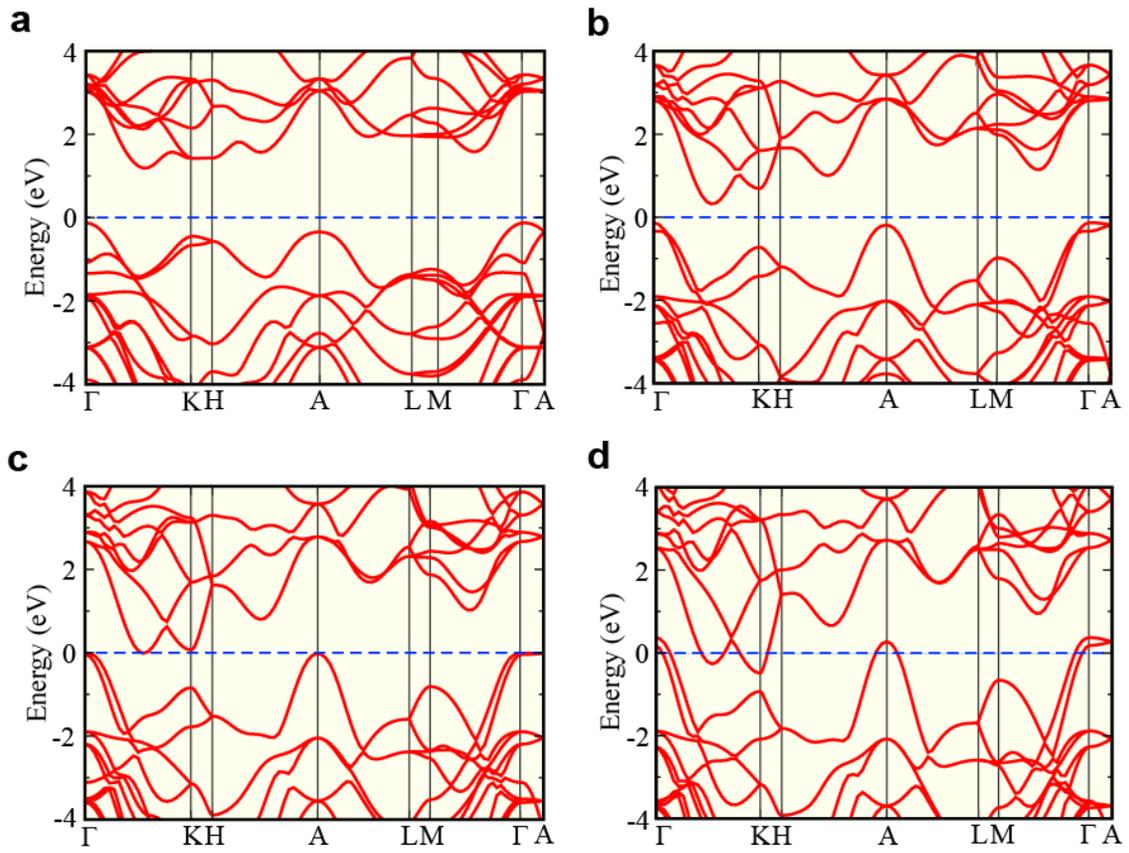

**Figure 5 | Calculated band structures of *2H$_c$*-MoSe$_2$.** (**a**) Ambient pressure, (**b**) 23 GPa, (**c**) 41 GPa, and (**d**) 58 GPa. Blue dotted line shows the Fermi level ($E_F$).

**Supplementary Information**

**Supplementary Figure 1**

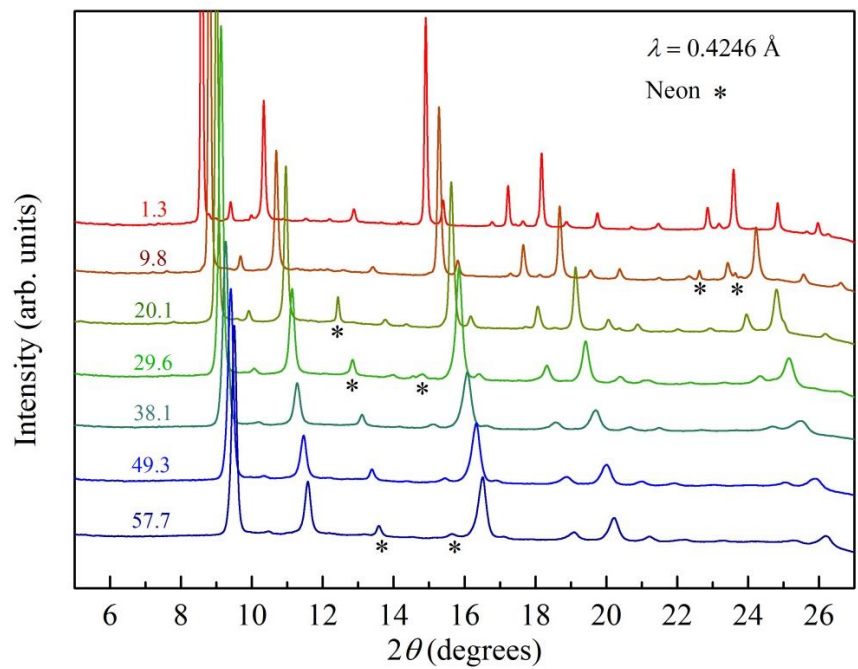

**Supplementary Figure 1.** Representative XRD patterns under pressure. Numbers represent pressures in unit of GPa. Asterisks indicate diffraction peaks from the pressure transmitting medium neon.

**Supplementary Figure 2**

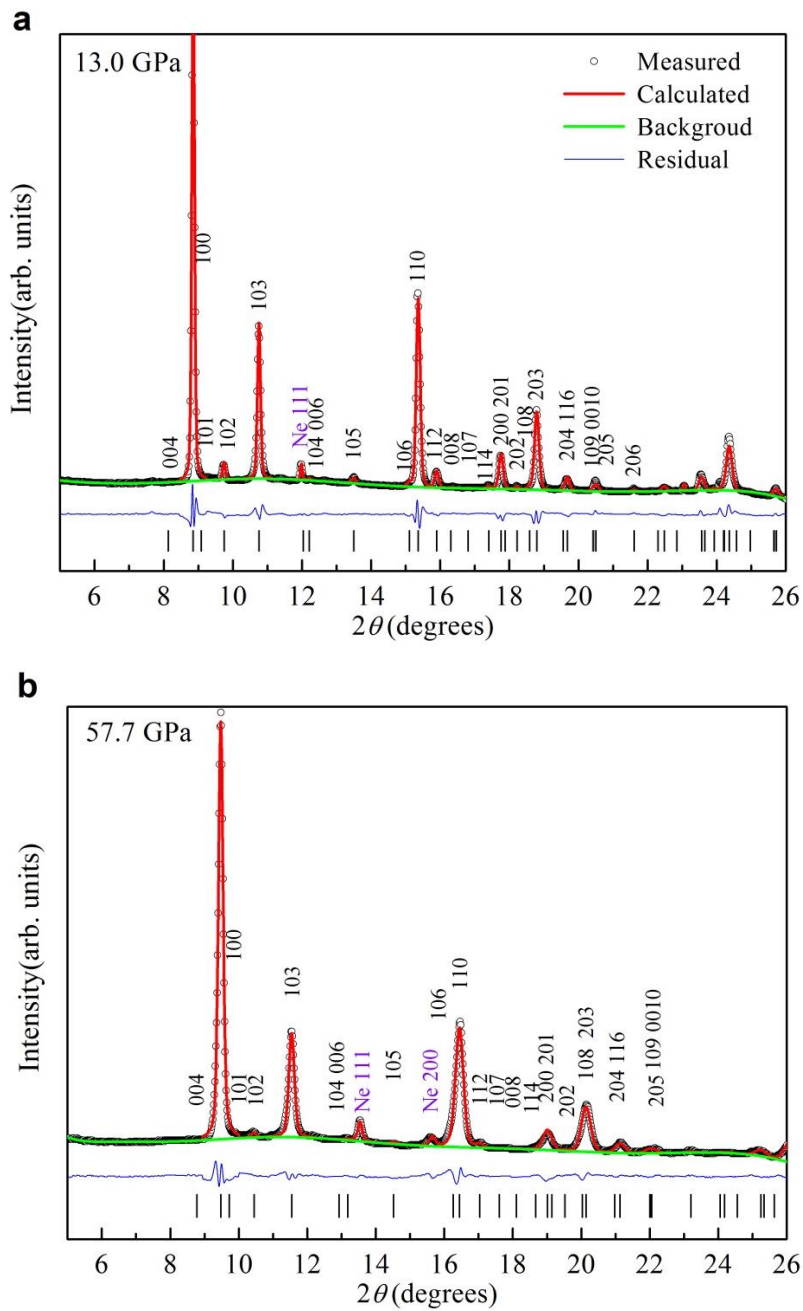

**Supplementary Figure 2.** Representative Rietveld refinement results for XRD data at (**a**) 13.0 GPa and (**b**) 57.7 GPa. The red lines and open circles represent the Rietveld fit and the measured counts respectively, and the blue lines give the residual intensities. The vertical bars indicate the predicted peak position of $MoSe_2$. Black labels show the diffraction peaks index of $MoSe_2$, and purple ones are from the pressure medium neon.

**Supplementary Figure 3**

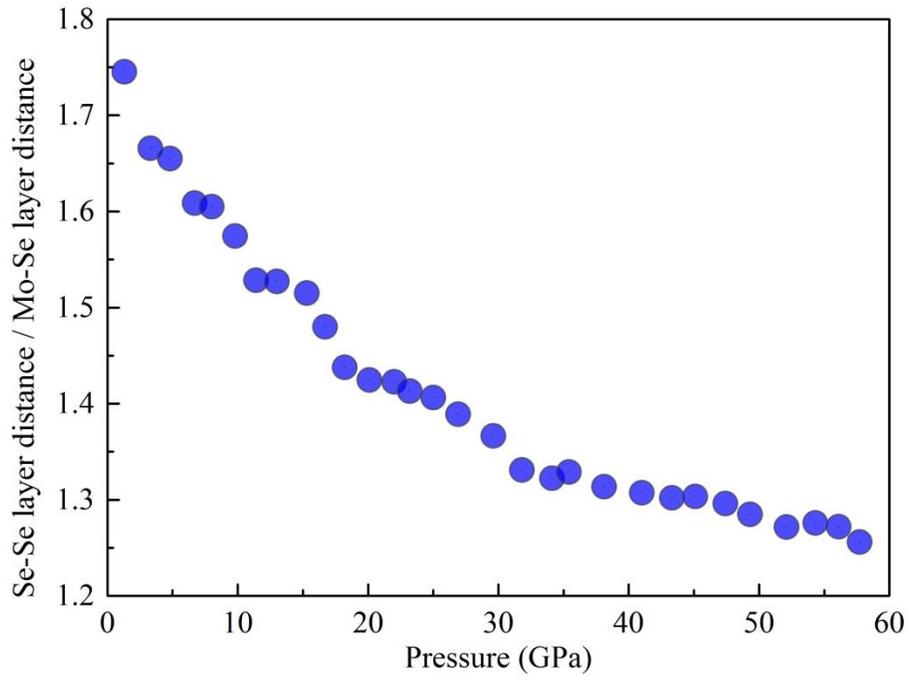

**Supplementary Figure 3.** Ratio of Se-Se layer distance to Mo-Se layer distance under pressure. This ratio is determined by the atomic positions from the Rietveld refinements.

**Supplementary Figure 4**

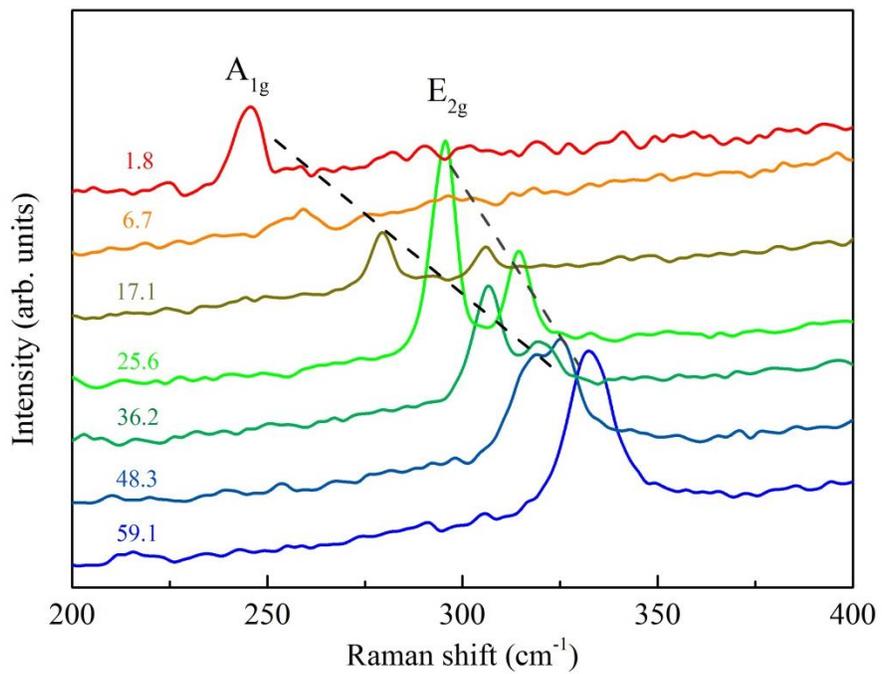

**Supplementary Figure 4.** Representative Raman spectra under pressure. Numbers represent pressures in unit of pressure.

**Supplementary Figure 5.**

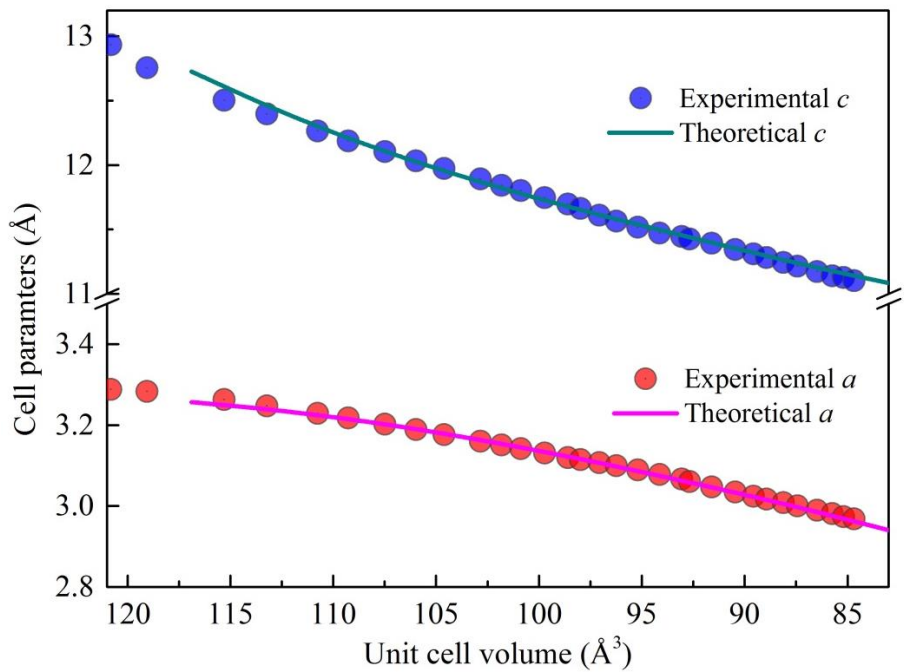

**Supplementary Figure 5.** Comparison of experimental and theoretical cell parameters at different unit cell volumes. Circles represent experimental data and lines are from calculations.

**Supplementary Figure 6**

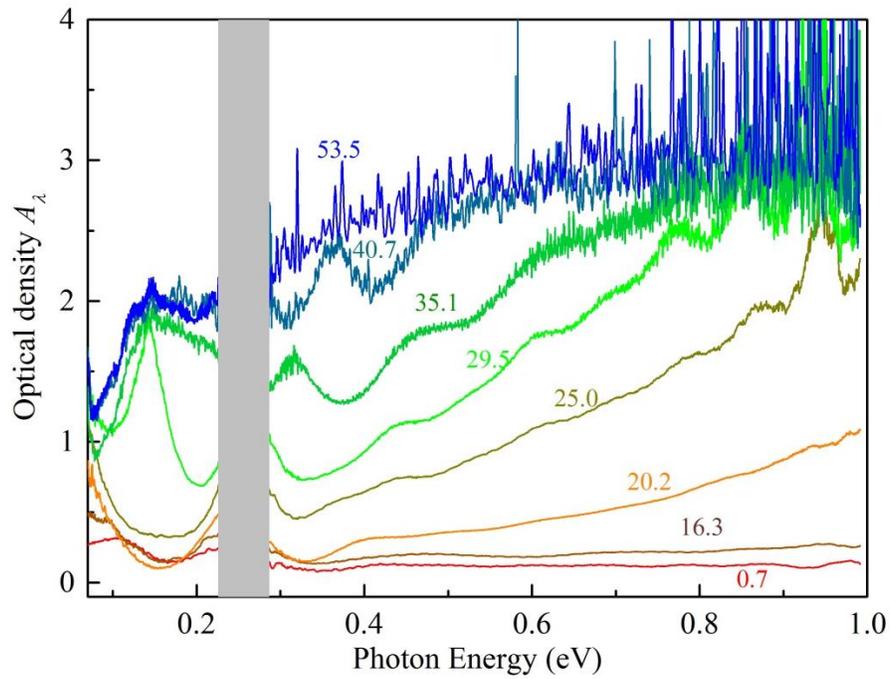

**Supplementary Figure 6.** Representative OD vs photon energy curves under pressure. The 0.23-0.28 eV region is obscured by diamond absorption. Numbers represent pressures in unit of GPa. The fluctuation of data (OD between 2 and 4) mainly results from the sharp decrease in transmittance under pressure, e.g. OD = 3 means transmittance = 0.001, which then makes the noise comparable to signal.

**Supplementary Figure 7**

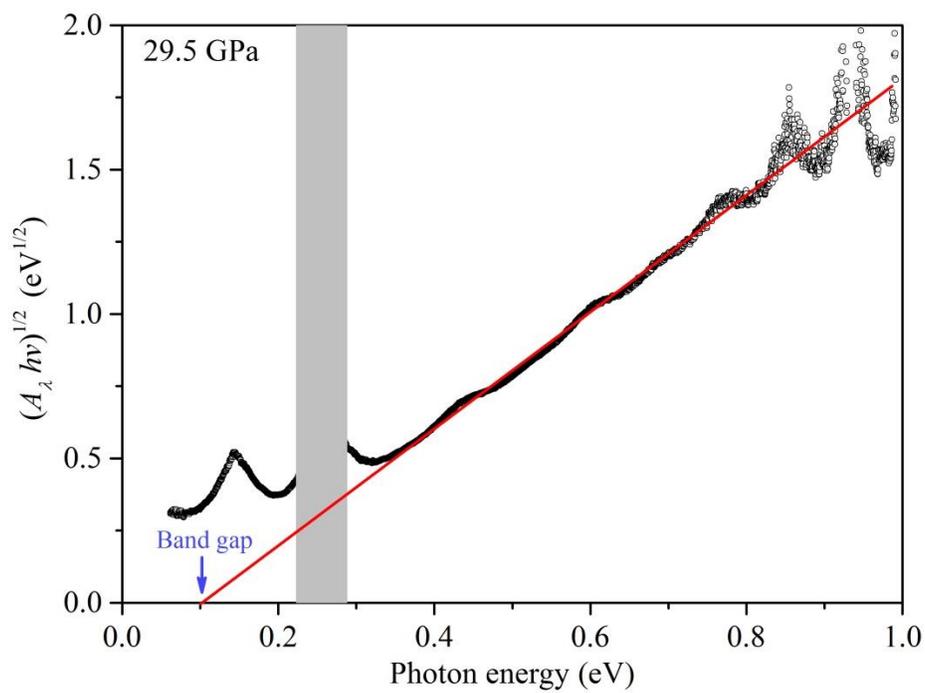

**Supplementary Figure 7.** Representative extrapolation for the indirect band gap at 29.5 GPa. The 0.23-0.28 eV region is obscured by diamond absorption.

**Supplementary Figure 8**

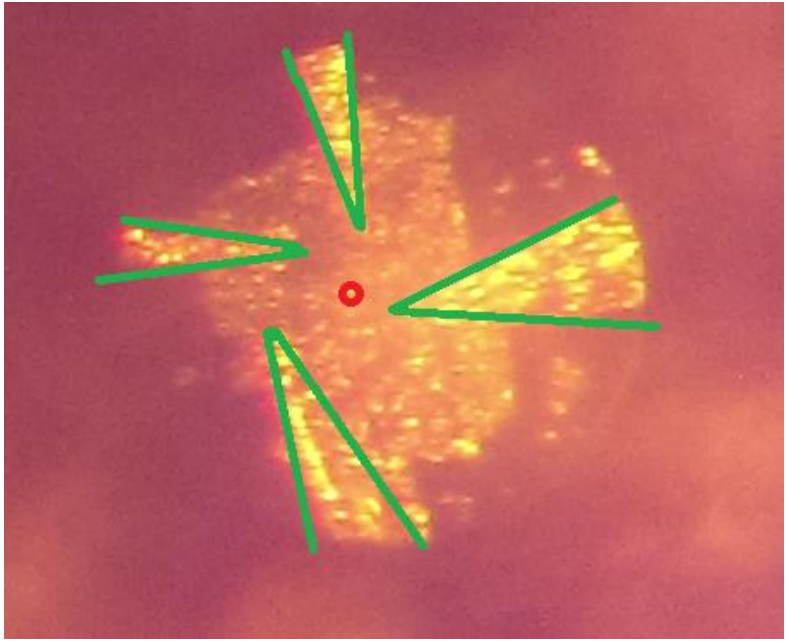

**Supplementary Figure 8.** Microphotography of MoSe$_2$ with four Pt electrodes attached to inside a DAC. The image is taken at ~ 30 GPa.

**Supplementary Figure 9**

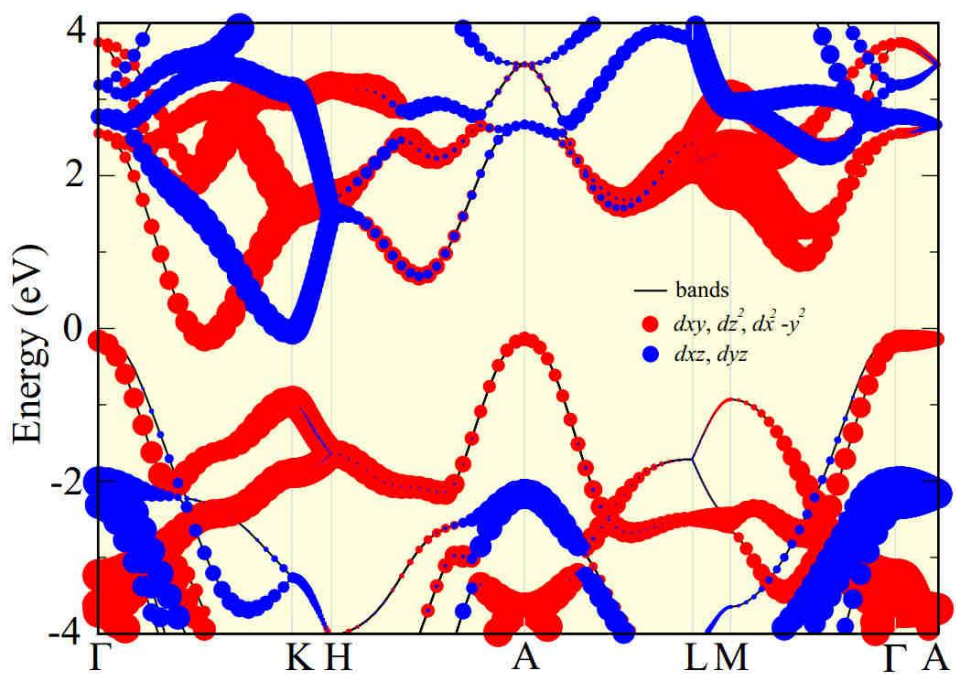

**Supplementary Figure 9.** Orbital information from band structure calculations at 41 GPa. The size of dot is proportional to the density of state of electrons.

**Supplementary Table S1.** Representative Rietveld refinement results from XRD data. Errors are given by GSAS-EXPGUI package.

| Pressure (GPa) | Space group | Volume (Å³) | Cell parameter | | Position parameter | | | |
|---|---|---|---|---|---|---|---|---|
| | | | $a$ (Å) | $c$ (Å) | Atom | x | y | z |
| 13 | $P6_3/mmc$ | 104.6(1) | 3.176(1) | 11.974(2) | Mo | 1/3 | 2/3 | 1/4 |
| | | | | | Se | 1/3 | 2/3 | 0.892(1) |
| 31.8 | $P6_3/mmc$ | 94.1(1) | 3.078(1) | 11.471(2) | Mo | 1/3 | 2/3 | 1/4 |
| | | | | | Se | 1/3 | 2/3 | 0.900(1) |
| 57.7 | $P6_3/mmc$ | 84.7(1) | 2.969(1) | 11.099(2) | Mo | 1/3 | 2/3 | 1/4 |
| | | | | | Se | 1/3 | 2/3 | 0.904(1) |

**Supplementary Note 1: X-ray diffraction measurements and analysis**

Synchrotron X-ray diffraction (XRD) data was measured in beamline 16-BMD, Advanced Photon Source (APS), Argonne National Laboratory (ANL). The sample-detector distance was set at 304.32 mm with X-ray wavelength at 0.4246 Å. Neon served as the pressure transmitting medium. MAR345 image plate was used to collect the diffraction data. The 2D diffraction data with diffraction rings was then integrated into 1D diffraction patterns through Fit2D program. During compression, all diffraction peaks continuously shift to higher $2\theta$ (smaller $d$-spacing) as seen from Supplementary Fig. 1. No new diffraction peak from $MoSe_2$ is seen throughout our measurements. Notice that diffraction peaks from neon (marked by asterisks) start to appear at 9.8 GPa. Decompression run shows that the shifts of all peaks are reversible.

We performed Rietveld refinement, a commonly used least-squares approach in solving powder XRD data, using the GSAS-EXPGUI package[1]. Representative Rietveld refinement results are shown in Supplementary Table.

In Supplementary Fig. 2, Rietveld refinement results for XRD data at 13.0 GPa and 57.7 GPa are shown. The red lines and open circles represent the Rietveld fit and the measured counts respectively, and the blue lines give the residual intensities. The vertical bars indicate the predicted peak position of $MoSe_2$. Black labels show the correspondent diffraction peaks index of $MoSe_2$, and purple ones are from the pressure medium neon. The $2H_c$-type structure can well fit all XRD patterns which supports the absence of structural transition.

The refined atomic positions then determine the ratio of Se-Mo layer distance to Se-Se layer distance. Supplementary Fig. 3 shows the ratio of Se-Mo layer distance to Se-Se layer distance, which drops fast at low pressure and but decreases much slower at high pressure. It indicates the gradual closure of the van der Waals (vdW) gap in between Se-Se layers.

**Supplementary Note 2:    Raman measurements**

To probe the change of phonon modes of $MoSe_2$ under pressure, we measured its Raman spectra at high pressure. The spectra were collected using a Renishaw inVia micro Raman system with a 514 nm laser excitation line, at Extreme Environments Laboratory, Stanford University. Neon served as the pressure transmitting medium. Silicon's 520.4 cm$^{-1}$ line was used for spectrum calibration before our measurements. To avoid overheating our sample, the laser intensity was set at 1 mW. The collection time was set at 60 s for each data point.

Representative Raman spectra during compression are shown in Supplementary Fig. 4. Two Raman modes $A_{1g}$ and $E_{2g}$ shift continuously under pressure. Each peak is fit by a lorentzian function using PEAKFIT. At above 57 GPa, these two peaks become so close that they could not separate by peak fitting. Notice that no new Raman mode is seen, which further supports the stability of the $2H_c$ structure. Decompression experiments show the shifts of these peaks are reversible.

**Supplementary Note 3: Theoretical and experimental unit cells.**

The Vienna *ab initio* Simulation Package (VASP)[2,3] was employed to optimize crystal structures and calculate electronic structures with the framework of local density approximation density functional theory[4]. The projector augmented wave (PAW)[5] pseudo-potential was used, and the kinetic energy cutoff was fixed to 450 eV for all the calculations.

Supplementary Fig. 5 shows the comparison of theoretical and experimental cell parameters. The theoretical $2H_c$ unit cell information is obtained with fixed volume and fully relaxed lattices and atoms. The experimental data is from the Rietveld refinements, as shown in Supplementary Note 1. Circles represent experimental data and lines are from calculations. Good agreements are found in between the theoretical and experimental data. It needs to be pointed out that the largest discrepancy is at large volume (low pressure), where the calculated *c* is slightly larger than the experimental value, and the calculated *a* being smaller. This results from the well-known inadequacies of *ab initio* calculations in describing the weak forces − in this case the vdW force (in between Se-Se layers).

**Supplementary Note 4: Infrared measurements and analysis**

High-pressure IR measurements were conducted in beamline U2A of the National Synchrotron Light Source (NSLS), Brookhaven National Laboratory (BNL). A MoSe$_2$ single crystal (thickness ~ 4 μm) was sandwiched between the pressure transmitting medium (KBr) and one side of the culet. Infrared microspectroscopy was performed on a Bruker Vertex 80v FT-IR spectrometer coupled to a Hyperion-2000 microscope with a MCT mid-band detector. Fringes in raw IR data were removed by filtering high frequency harmonic after Fourier transformation. Supplementary Fig. 6 shows Representative optical density (OD) vs photon energy curves under pressure. OD is defined as –log(*T*), *T* is transmittance. There is no sharp cut-offs in these curves, which supports the "indirect" feature in the electronic structure. The band gap closure is seen from gradual lifting of these curves.

For an indirect-band-gap semiconductor, the absorption coefficient is proportional to the square of the photon energy and band gap. Using this empirical model for semiconductors, we obtained the indirect-band-gap $E_g$ via linear extrapolations of $(hvA_\lambda)^{1/2}$. A representative fitting at 29.5 GPa is shown as result Supplementary Fig. 7.

**Supplementary Note 5:   Electrical resistivity measurements**

For temperature-dependent four-probe resistivity measurement, cubic BN was used as the insulating layer and pressure transmitting medium. A single crystal of $MoSe_2$ with suitable size was chosen for measurement. Ruby was used as the pressure calibrant. The four electrodes with sharp heads were cut from Pt foils, see Supplementary Fig. 8. The Van der Pauw geometry of these four electrodes is outlined by green lines. The position of ruby is marked by red circle. The temperature-dependent sheet resistance of the sample was measured by cooling down to 10 K in a liquid helium cryostat after changing each pressure at room temperature. The difference in pressures before and after the cooling and warming cycle is typically ~ 5 %.

**Supplementary Note 6:    Orbital details in band structure**

HSE06 hybrid function[6] was employed for calculations. The **k**-points mesh is taken as 12 × 12 × 10 for all bulk self-consistent calculations. The conduction bands and valence bands near the $E_F$ show large movements. At ambient pressure, the *dxz* and *dyz* dominated conduction bands are further away from the Fermi level than the *dxy* and $dx^2 - y^2$ dominated bands. Interestingly, two band minima are observed at high pressure. For example, at 41 GPa as seen in Supplementary Fig. 9, one *dxz* and *dyz* dominated conduction band quickly goes down at K point to form two conduction band minimum. This is because that *dxz* and *dyz* orbitals gain more overlap with Se *p* orbitals to widen the band dispersion than *dxy* and $dx^2-y^2$ within the $2H_c$ structure under pressure. The conduction band minimums may play a role in determining the transport properties.